# IoT-Enabled Hemodynamic Surveillance System: AD8232 Bioelectric Signal Processing with ESP32

Samson Jebakumar S [1,a)] Hemalatha R J [2,b)] Shubham Malhotra [3,c)]
Shivapanchakshari T G [4,d)] Lokesh K [5,e)] Dev Anand D [6,f)]

Author Affiliations

[1] *Department of Biomedical Engineering, Vels Institute of Science Technology and Advanced Studies, Chennai, Tamil Nadu, India.*

[2] *Department of Biomedical Engineering, Vels Institute of Science Technology and Advanced Studies, Chennai, Tamil Nadu, India.*

[3] *Department of Software Engineering, Rochester Institute of Technology, Rochester, NY, USA, 14623.*

[4] *Department of Electronics and Communication Engineering, Cambridge Institute of Technology, Bengaluru, Karnataka, India.*

[5] *Department of Computer Science and Engineering, Gojan School of Business and Technology, Chennai, Tamil Nadu, India.*

[6] *Department of Computer Science and Engineering, Gojan School of Business and Technology, Chennai, Tamil Nadu, India.*

Author Emails

[f)]Corresponding author: *org.vipstech@gmail.com*
[a)] *ssamsonjebakumar@gmail.com*
[b)] *hemalatharj.se@vistas.ac.in*
[c)] *shubham.malhotra28@gmail.com*
[d)] *hod.ece@cambridge.edu.in*
[e)] *lokeshkcse314@gmail.com*

**Abstract:** This dissertation proposes an electrocardiogram (ECG) tracking device that diagnoses cardiopulmonary problems using the Internet of Things (IoT) desired results. The initiative is built on the internet observing an electrocardiogram with the AD8232 heart rhythm sensor and the ESP32 expansion kit, using an on-premise connected device platform to transform sensing input into meaningful data. That subsequently supervises an ECG signal and delivers it to an intelligent phone via Wi-Fi for data analysis. That is the pace of the circulating. Assessing body temperature, pulse rate, and coronary arteries are vital measures to defend your health. The heartbeat rate may be measured in two ways: there are by palpating the pulse at the wrist or neck directly or other alternative by utilizing a cardiac sensor. Monitoring alcohol levels in cardiac patients is critical for measuring the influence of liquor on their health and the efficacy of therapy. It assists in recognizing the association between alcohol consumption and cardiac issues, rather than rhythm recorded in beats per minute (bpm). An IR transmitter/receiver pair (OLED) needs to stay compatible up near the sensor's knuckle current or voltage pulse. The detector's electrical output is evaluated by suitable electronic circuits to produce a visual clue (digital display). We must design a cost-effective, user-friendly, and efficient ECG monitoring system with contemporary technology for both persons imprisoned by disease or aging, as well as healthcare professionals. Microcontroller combined with software. A smartphone application is created to monitor the cardiovascular health of distant patients in real-time.

## INTRODUCTION

Health expert systems are flourishing, notably in convertible and sophisticated nursing assets. Everyday living. Medical applications have tremendous importance in this swiftly growing and aggressive economy. Any shift in heart rate or rhythm, or fluctuation in the morphological shape of the ECG signal, is symptomatic of an arrhythmia, which may be diagnosed by the evaluation of archived ECG data [1].

Cardiovascular diseases (CVDs) are deemed as the most prevalent worldwide causes of mortality (WHO). Bangladesh has one of the greatest cases of cardiovascular disease (CVD) among developing countries, with 99.6% of the male population and 97.9% of the female population having at least one identified CVD risk factor.

Nevertheless, the healthcare system lacks adequate capacity to recognize and cope with the consequences of these deadly disorders. Constant monitoring of symptoms is necessary for persons with cardiovascular disease [2].

The recommended e-health system obtains ECG information regarding the patient's remote location utilizing healthcare monitors on consumer gadgets and a database of the controllers. The equipment includes four vital elements: patient route that is estimated, ECG signal gathering, patient record regular consumption, and hospital alert monitoring. The trajectory forecasting applies HMM to determine the subsequent position/location of the victim's body depending on his/her past locations. Subsequently, the cardiac readings acquired from the medical care detectors are transferred to the gadgets via Bluetooth for medical evaluation of patients with CVD. Further, depending on the expected future position as well as analysis of the cardiac indications, the system develops new sending instructions for healthcare operators. Eventually, the medical professional following the incoming guidelines from individual CVD patients applies the associated monitoring record to deliver proper therapy as necessary, particularly in instances of emergency [3].

Cloud computing represents some of the most attractive and exciting research pathways. This computing technique delivers both software and infrastructure services for consumers as well as services that are requested by clients through internet access. Considering the remarkable rise of cloud computing, the quantity of customers and requests is expanding fast. Therefore, enhancing the speed & accuracy of cloud computing is critically significant [4].

In recent years, various approaches have been stipulated to design effective monitoring of patient systems. These involve setting up ECG sensors to constantly capture and store ECG data on servers in the cloud for inspection, complementing cardiovascular sensors for measuring BPM and sensors that detect temperature for continuous temperature monitoring. Cloud computing allows remote storing and gaining access to the internet for parties. Some systems further enable a one-touch call option for emergencies. These enhancements strive to promote continual surveillance and accessibility for patients as well as caregivers [5].

The Internet of Things (IoT) is transforming into a widely dispersed, varying architecture that can consistently adapt to satisfy the needs of multiple enterprises and people. Fast innovations in IT-based technologies, notably cloud computing and the Internet of Things, have permitted public health monitoring, providing affordable medical treatment and aid, and seamless holistic management. As a result, integrating IoT with medical care is gaining growing interest from intellectual and industrial circles [6].

As technologies progresses quickly, there has been an increasing interest in creating adaptable and cheap ECG monitoring devices that may be used both in medical settings and by people at home. Smart healthcare monitor systems detect the wellness status, i.e., the rate of the pulse, the body temperature, oxygen consumption, glucose levels, the position of the body, ECG, EEG, and other things by applying sensors. The sensors are connected and handled by several microcontroller-based systems including Arduino, Raspberry Pi, etc. The microcontroller collects the data via a detector. The collected biological data is frequently kept on servers [7], [8].

This work solves the problems by constructing a heart disease monitoring system employing wireless sensors. The system monitors the heartbeat and aberrant heartbeat using linked sensors and utilizes the growing acceptance of IoT in the monitoring of patients remotely. The utilized sensor constantly checks the patient's heartbeat and evaluates the data to detect serious conditions and warn the physicians. The patient records, including the body temperature and heartbeat, is transferred to the cloud utilizing Wi-Fi. It lets the doctor to monitor in online, and it is an affordable, portable monitoring gadget, which additionally helps the patient heal quickly at their home. It also lowers the visits of healthcare facilities and gives an integrated approach for managing health easily.

## LITERATURE REVIEW

In this study, we apply an openMSP430 processor to develop a programmable system of components for analyzing signals and then implement it in compact monitoring equipment. The configuration of the system on the openMSP430-based infrastructure may be modified by programs for software, including the parameters of the system, filtering coefficient, as well as output format to accommodate various kinds of applications. We also integrate the wearable sensors with the data stream interpretation architecture to provide real-time data analysis and recognition. The installation outcomes indicate a 10 times reduction in time to response [9].

This research effort provides a cheap, portable ECG wireless device with feature extraction and a cardiovascular disease diagnosis method. The system design comprises an accessible ECG signal-generating circuit, a data transmission device, and an effective device. Someone may quickly check the probability of any heart problem with this approach. The benefits of this system could be beneficial before, during, and after a cardiovascular arrest for live-time tracking of a patient at any site. It might also minimize dying due to cardiac attack and other coronary disorders and, more particularly, offer health care by expert doctors to remote regions. This suggested study is more advantageous for health security with minimal cost [10].

This paper summarizes the implementation of IoT in health-tracking systems. While IoT is being utilized in many domains of health science, there are chances for future advancement as well as research. Timely identification of health issues may enable the person to implement critical emergency interventions, perhaps preserving the patient's life. The Internet of Things may facilitate this matter. IoT-related health monitoring systems can follow patients throughout real periods and notify them of any anomalies. However, the design of the IoT must include mechanisms to ensure the effective protection of sensitive information. The deployed sensors need to be compact to facilitate their integration into several systems. As a result, the integration of various ML and DL methodologies might enhance the efficiency and resilience of the systems. The premise of an advanced healthcare surveillance system utilizing Internet of Things frameworks is an innovative development in medical research, aimed at reducing health complications and preventable fatalities [11].

The provided technology proved effective in absorbing and interpreting the massive data produced by continual observation of patients in real-time. Effective medical facilities may apply the suggested structure to monitor their patients at their residences in an instant. The suggested approach exploits the enormous capacity of the clouds for both data storage & processing. Furthermore, a localized component that observes patients with equivalent efficiency in an instance of internet disruption or malfunction in the cloud-based infrastructure has been developed. Our investigation has enabled us to conclude that the recommended model is rapid, accurate, and tolerant of faults whenever it applies to monitoring patients with BP concerns and forecasting their medical condition with high accuracy and minimal false alerts. Overall, this study demonstrates that combining sampling procedures in parallel via Spark increases the accuracy of classification and reduces the total erroneous rate, particularly for minority categorization (emergency class). The recommended technique exhibited greater efficiency, enhanced precision, and an improved F-measure than previously suggested systems that utilized Hadoop as well as association policies without controlling imbalanced datasets [12].

In this study, we created an inexpensive ECG system utilizing sophisticated IoT methods for continuous, long-lasting tracking. The IoT-assisted electrocardiogram system design delivers results equivalent to traditional systems. The ECG signal is obtained using an AD8232 ECG monitoring with AgCl gel electrode as well as a Node MCU microcontroller with inside-the-chip Wi-Fi. Data is transferred to the Internet of Things cloud for long-term retention and extensive coverage. Algorithms calculate essential ECG characteristics and identify arrhythmia. An online and mobile application GUI visualizes the electrocardiogram (ECG) signals, parameters, and heart rhythm state. The system combines cheap cost, excellent mobility, and continuous tracking with arrhythmia detection [13].

The work is a more sophisticated version of present patients health monitoring systems since it incorporates IoT technology, enabling more rapid and safe data transfer. IoT connects patients, physicians, equipment, and software into one network, making it easy. It saves time for physicians as well as patients and is particularly useful for persons in remote regions without means for regular hospital visits. It is effective for physicians, allowing speedier diagnosis. The system employs an STM32F429 microprocessor, unlike most current systems that use the PIC microcontroller. This microcontroller has minimal power consumption and fits Embedded Systems applications [14].

In this research, we provide a smartphone-based ECG monitor exploiting an ESP32 microcontroller and AD8232 sensor. ECG data arrives at a smartphone application via Bluetooth to detect heart arrhythmias. Compared with earlier technologies, it provides expenses, accessibility, and brevity while adopting other health devices for total wellness surveillance. However, it has limits in both accuracy and reliability for individuals with increased risk. Future research may target noise reduction filters to increase system accuracy and dependability [15].

This research indicates that technologies such as IoT and microcontrollers have proved beneficial for health surveillance, particularly in providing instantaneously advantages for ECG heart activity. These systems are affordable, adaptable, and simple to use, providing significant benefits to individuals since they may operate in distant areas and facilitate online health data storage. It would be simple for the doctors to review the patient health data and reply swiftly. Some technologies even forecast the health concerns using clever algorithms. These systems would be more precise and secure and intelligent with the superior technology.

## PROPOSED METHODOLOGY

The allegedly likely ECG system is geared towards fabricating ECG signals at any given period practically anywhere as depicted in "Figure 1". The AD8232 records the ECG electrical signals in real-time, and then it appraises the fluctuations obtained by the root activity and rhythm and then the other individual's detectors, like the body temperature sensor, which is used to evaluate the body temperature of the patient, and then the alcohol sensor, which is used to monitor the patient pulse rate per minute. Now all the sensor-captured understanding, such as ECG sensor data, temperature sensor data, alcohol level data, and pulse data, passes on to the ESP32 microcontroller. Now the ESP32 examines every bit of data that is prepared for dissemination, and then it also checks to reveal if there are any anomalies, such as erratic heartbeats or high body temperature. If the person drinks alcohol, it will activate notifications.

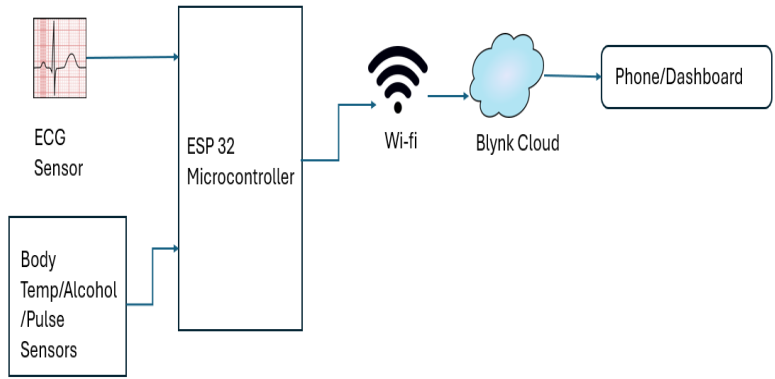

**Figure 1:** *Proposed System Diagram.*

The ESP32 capabilities an onboard Wi-Fi module that engages the information it has extracted to the Blynk web system. The Blynk cloud gives a live display that exhibits the ECG signals. ECG signals, temperature of the body, alcohol stage, and heart rate for each person Now medical professionals and parents or caregivers, as well as inevitably the patient, can consult their Blynk dashboard on their cellphones or computers. These folks may access their real-time information; a practitioner may also analyze the consumer's medical state and take action. The AMOLED is coupled to the ESP32, and it displays the ECG signal and any critical data if any odd activities, such as tachycardia or growing alcohol level, occur; the alert is activated to warn patients and caregivers.

"Figure 2" depicts the circuit diagram of a coronary heart illness diagnostic system. It leverages the ESP32 microcontroller. It encompasses the sensors, such as the ECG sensor and body temperature sensor, for observing the heartbeat. It demonstrates it is evaluated; that information is sent by GSM for alarms and Wi-Fi for storage in the clouds, observing via the internet. The LM256 buck converter supplies the power, and then the OLED shows a live indicator of return. It is an inexpensive and easily available solution that enables great, remote-based management of health. It is a major benefit for the medical providers.

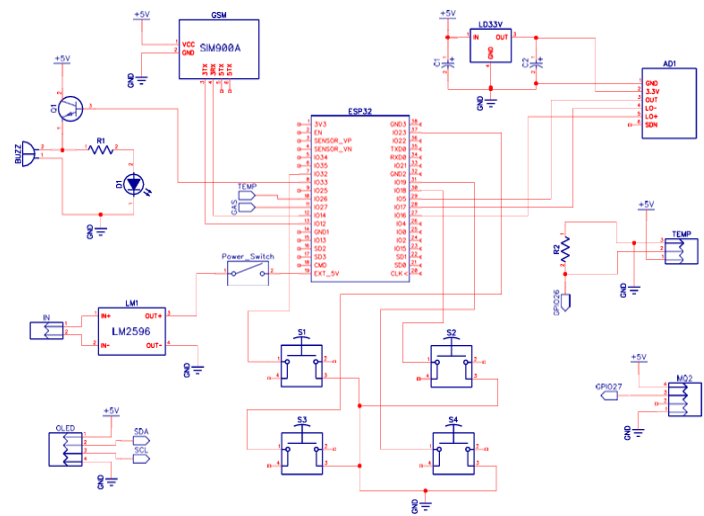

**Figure 2:** *Proposed Circuit Diagram.*

## Description Components

The architecture and physical frameworks of a wireless Internet of Things-associated environmental surveillance network are described in this section. Materials and methods, design concepts, and design specifications are a few of the tasks completed.

### *ESP32 Microcontroller*

The ESP32 microcontroller device is especially prominent in the area of IoT since it offers outstanding capabilities and involves power effectively. It was created by Espress if Systems and is incredible since it contains built-in Wi-Fi and Bluetooth, complemented with a dual-core CPU. It lends it an appealing alternative for tasks entail communication and control. It works well in several builders and languages for programming, so it's uncomplicated to utilize with modern systems. It also requires exceptionally tiny amounts of power, which makes it appropriate for battery-powered gadgets. It's a relatively cheap and cost-effective choice for a broad variety of IoT applications. In summary, the ESP32 is a strong microcontroller that's employed in numerous distinct IoT applications [17].

### *AD8232*

The Heartbeat Rate Monitoring Kit including AD8232 ECG sensor module Kit for Arduino is an affordable board employed to monitor the electrical function of the heart. The flow of electricity may be recorded as an ECG reading, or an electrocardiogram, as well as illustrated as an analog response. ECGs may be quite raucous; the AD8232 Single Leads Electrical Heart Rate Analyzer operates as an amplifier operation to permit the flawless extraction of a distinct signal from the PR as well as QT intervals. The AD8232 ECG component is a consolidated signal processing unit for ECG and biopotential measuring tasks [5], [18].

The ECG Components AD8232 Heart ECG Tracking Sensor Module Kit for Arduino is engineered to gather, amplifying, & evaluate small bio-potential data during the occurrence of noisy situations, especially those produced by circulation or remote electrode placements. Its AD8232 Heart Rate Monitor splits out 9 links within the IC which you may solder pins, wires, or other connections onto. SDN, LO+, LO-, OUTPUT, 3.3V, & GND provide crucial pins for controlling the track with a single Arduino or another programming board.

The (ECG) waveform captures the electrical signals of the coronary artery. The P wave represents cardiac depolarization (an atrial contraction), whereas the complex of the QRS indicates a depolarization of the ventricular muscle (ventricular contraction). The appearance of a T wave represents the repolarization of the ventricular muscles, indicating readiness for the subsequent contraction. Critical intervals such as PR and QT assess the time of electricity conduction. The electrocardiogram (ECG) is an essential diagnostic instrument for identifying cardiac disorders, and correct electrode positioning guarantees precise measurements.

### *Gas Sensor*

The gas sensor module is an electrical device that detects and identifies various kinds of gases. They are often used for identifying dangerous or volatile gases and monitoring concentrations of gases. Gas detectors are deployed inside factories and industrial organizations to discover gas leakages and to monitor smokes and carbon monoxide levels in residences. Gas monitors varies significantly regarding size (portable and fixed), ranges, as well as sensor capabilities. They are generally part of an entire integrated system, including biohazard and monitoring systems, and their functions typically correspond to a noticeable alert or the interface [8].

### *OLED Display*

The organic light-emitting diode, generally referred to as OLED, is a solid-state system that comprises thin layers of natural compounds that create outstanding light on the application of electricity. They are manufactured via a succession of natural thin sheets put among a pair of conducts. Organic layers included in an OLED are superior to the crystallographic levels in LEDs or LCDs since they are more adaptable, less intense, and thinner than crystalline levels. Because of the existence of natural levels, OLEDs are lighter than LEDs. OLEDs require fewer resources than LCDs and LEDs. They feature quicker reaction periods and greater image excellence and observation ranges than LEDs.

*Temperature Sensor*

The DHT11 is an affordable digital sensor used for measuring both humidity and temperature. This sensor may be rapidly coupled to an appropriate microcontroller, including the Raspberry Pi, Arduino, etc., to monitor both temperatures and humidity rapidly. The DHT11, both a temperature and humidity sensor, is provided as a sensor as well as the module itself. The disparity between this sensor & the component is the pull-up resistance & a power-on LED. DHT11 serves as an approximate humidity monitor. To monitor the ambient environment, this type of sensor that employs a thermistor as well as a capacitive humidity sensor. The sensor DHT11 contains a capacitance humidity-detecting circuit as well as a thermistor for sensing temperatures. The humidity-detecting capacitors have a pair of electrodes that have a moisture-retaining substrate as a separator among them [19].

*Buzzer*

It is a sound-signaling gadget including a beeper or buzzer that might be of the electromechanical, piezoelectric, or mechanical type. The fundamental aim of this process is to change the signals from audio files to sound. On the whole, it is fed by a direct current and then utilized in timepieces, concern stuff, apparatus for printing alerts, computers, etc. Depending on the diverse plans, it may generate numerous noises, including an alert, tunes, an alarm bell, & a siren. The layout of the pins of the buzzer is provided below the figure. It contains two pins, including plus & minus. The plus terminal of this is marked with the '+' symbol or a longer terminal. The corresponding terminal is supplied with 6 volts, while the negative terminal is designated with the '-' symbol or shorter terminal, and its pin is attached to the GND terminal [16].

## Implementation of Proposed System

*Hardware Implementation*

The ECG is among the most helpful studies in cardiology. Electrodes attached to the chest and/or limbs record minor voltage variations as potential variations, which then translate into a graphical trace. The ECG is among the more popular monitoring tests conducted on individuals. The connection of electronic elements allows effective observation of the heart.

The ESP32 microcontroller is utilized as the central control mechanism in the illustration. Many components are interconnected, primarily the gas sensor, temperature sensor, ECG sensor, buck converter, buzzer, and OLED panel. These modules, utilized with the ESP32, comprise a complex network of sensors for data collecting and analysis. The ESP32 examines the received information and initiates suitable reactions, such as issuing alarms or notifying the user about the progress via the module. This record will offer a full review of sinus rhythm, including its structure, features, normal electrical activity, ECG determination, variances, medical importance, therapy concerns, and closing with vital points and associated visuals. Representation of an ECG demonstrating a straight and established regular sinus rhythm. The ECG exhibits prominent P waves, conventional QRS complexes, and T waves, suggesting the healthy electrical function of the heart. "Figure 3" shows the hardware integration.

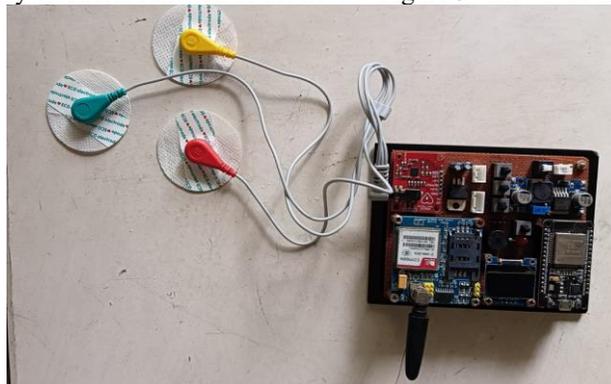

**Figure 3:** *Proposed Hardware Implementation.*

*Software Implementation*

The AD8232 One Lead Cardiac Rate Monitoring serves as an operational amplifier that assists in detecting clean signals from the PR as well as QT periods conveniently. The ECG modules AD8232 heart ECG tracking detector circuit is an integrating signals training blocks for electrocardiograms as well as bio-potential measuring possibilities. The ECG Module AD8232 Heart ECG Monitoring Sensor Module Kit for an Arduino is meant to capture, amplify, & analyze tiny bio-potential pulses in the context of boisterous situations, including those generated by movement or distant electrode location. The AD8232 Cardiac Rhythm Monitor takes off nine links within the IC which you may connect pins, wires, or other components onto. SDN, LO+, LO-, OUTPUT, 3.3V. "Figure 4" shows the software integration.

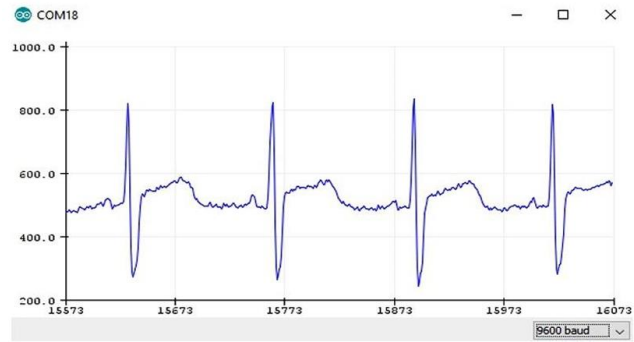

**Figure 4:** *Proposed Software Implementation.*

# RESULTS AND DISCUSSION

The IoT-based ECG system is built according to the suggested structure in the subsection Design of the proposed IoT-oriented Electrocardiogram Systems. The hardware of the method is shown in "Figure 5", is a device composed of AD8232, the electrocardiogram sensors, and the Node MCU microcontroller that is executed on an ESP32 and might be established in all everyday usage items like a belt, t-shirts, wristbands, car steering, office chairs, etc., thereby rendering the entire system accessible as well as perfect for a structure that is to be utilized for a constant, long period of ECG tracking. The ECG indicator is gained from the electrical activity of the human physique using the traditional AgCl gel electrodes with the AD8232 ECG sensor, after which the analog information is delivered sequentially to the Node MCU microcontroller, which gets it on the analogous pin A0 as well as employing a 10-bit built-in ADC, transforms the analog information into digital formats for more analyzing. The digitized information is then prepared based on a system implemented in the Node MCU microcontroller; the processed ECG information is then transferred to the Blynk IoT cloud employing Wi-Fi. "Figure 6" shows the testing image for hardware integration result.

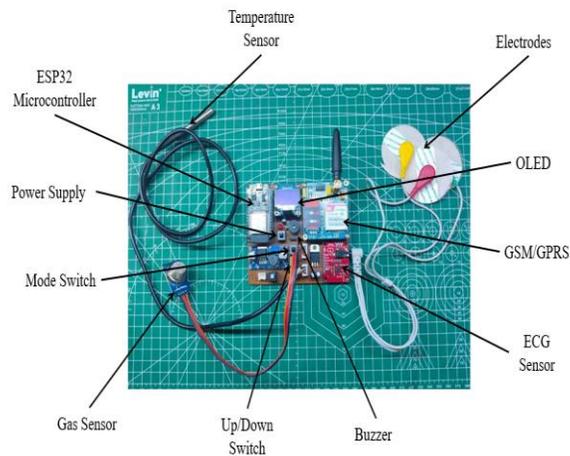

**Figure 5:** *Proposed Hardware implementation with Components.*

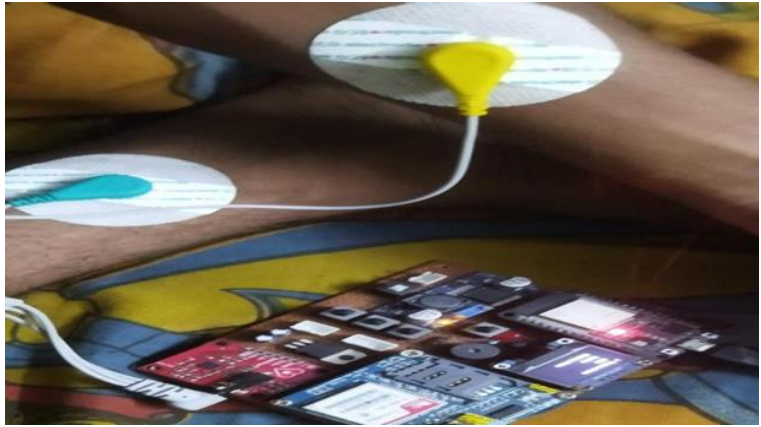

**Figure 6:** *Testing Image for Proposed System.*

"Figure 7" depicts the ECG signal for the heart rate. It produces waveforms with regularity that correspond to the electrical activity of the heart. The signal is vital for monitoring coronary heart diseases and recognizing the faults in the overall plan that is delivered. Study this ECG data is gathered using sensors and sent via IOT-based devices for real-time early medical identification. It is easy for both, patients and physicians, to gain online access via the Blyk Cloud web server.

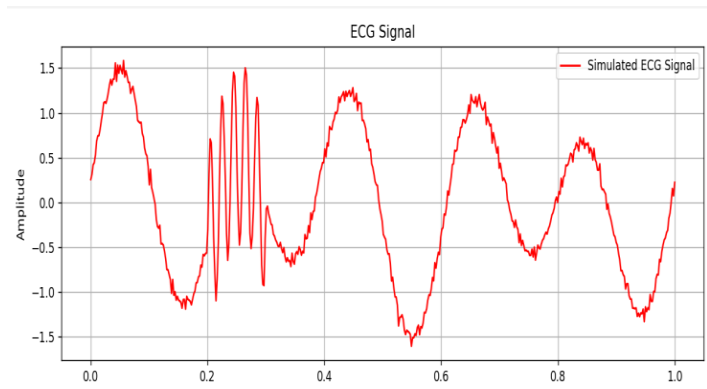

**Figure 7:** *ECG Signal.*

"Figure 8" depicts the ECG signal in a red curve with the peaked time noted as blue dots. The peak timings are highlighted above for each peak, indicating crucial places in the waveform. The IoT-based technologies allow live-time tracking and automate the identification of these elevations for cardiac study. This method determines the heart rate and variations correctly.

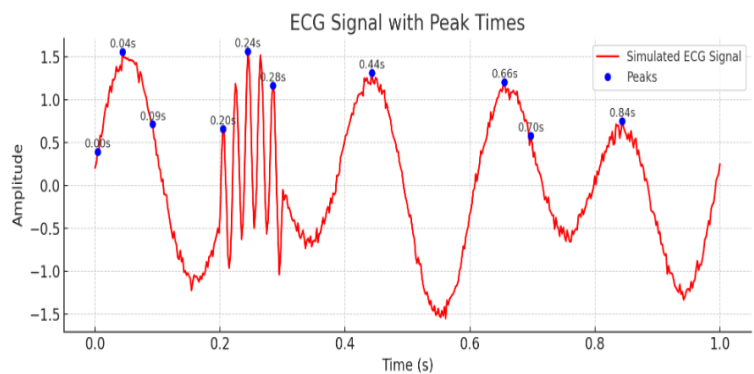

**Figure 8:** *ECG Signal wth Peak Time.*

# CONCLUSION AND FUTURE SCOPE

The effect of this study offers an in-depth analysis concerning the recommended approach, which is a patient-tracking technology. This study analyzes different methods that may be executed for giving the relevant information about patients that are physiological issues through considering unique body factors into consideration. The results obtained can consequently be incredibly helpful when evaluating one's development of the many physiological parameters of the patient separate from the therapy settings. The outcomes supply proof that the consumption of modern cutting-edge technical gear lets speedy reliable observation of patients with unusual characteristics. The combination of these improvements and frequent treatment by adequately sought-after specialists is a low expense and leads to gains and recuperation. Future research is focused on combining machines and IoT for increased ECG tracking, accuracy in robotic-assisted operations, and tailored medical care adapted to individual requirements for better standards of life.